\documentclass{aa} 

\usepackage{graphicx} 

\begin{document}

   \title{The oxygen abundance distribution in M101}

   \author{ L.S. Pilyugin }

   \institute{   Main Astronomical Observatory
                 of National Academy of Sciences of Ukraine,
                 Goloseevo, 03680 Kiev-127, Ukraine, \\
                 (pilyugin@mao.kiev.ua)
                 }

  \offprints{L.S. Pilyugin }

\date{Received 15 March 2001 / accepted 00 Month 2001}

\abstract{The well-observed spiral galaxy M101 was considered. The radial 
distributions 
of oxygen abundances determined in three different ways (with the classic 
T$_{e}$ -- method, with the R$_{23}$ -- method, and with the P -- method) were
compared. It was found that the parameters (the central oxygen abundance and the 
gradient) of the radial (O/H)$_{P}$ abundance distribution are close to 
those of the (O/H)$_{T_{e}}$ abundance distribution. The parameters of the 
(O/H)$_{R_{23}}$ abundance distribution differ significantly from those of the
(O/H)$_{T_{e}}$ abundance distribution: the central (O/H)$_{R_{23}}$ oxygen 
abundance is higher by around 0.4dex and the gradient is steeper by a factor
of around 1.5 as compared to those values in the (O/H)$_{T_{e}}$ abundance 
distribution.
The dispersion in (O/H)$_{P}$ abundance at fixed radius is rather small,
$\sim$ 0.08 dex, and is equal to that in (O/H)$_{T_{e}}$ abundance.
The dispersion in (O/H)$_{R_{23}}$ abundance at fixed radius is appreciably
larger, $\sim$ 0.16 dex, compared to that in (O/H)$_{T_{e}}$ abundance. 
It has been shown that the extra dispersion in (O/H)$_{R_{23}}$ abundances
is an artifact and reflects scatter in excitation parameter P at fixed radius.
}

\titlerunning{The oxygen abundance distribution in M101}

\authorrunning{Pilyugin L.S.}  

\maketitle

\section{Introduction}

By now spectra have been obtained for hundreds of H\,{\sc ii} regions in spiral 
galaxies. Accurate oxygen abundances can be derived from measurement of 
temperature-sensitive line ratios, such as [OIII]4959,5007/[OIII]4363. This 
method will be referred to as the T$_{e}$ - method. Unfortunately, in 
oxygen-rich H\,{\sc ii} regions the temperature-sensitive lines such as [OIII]4363 are 
too weak to be detected. For such H\,{\sc ii} regions, empirical abundance indicators 
based on more readily observable lines were suggested (Pagel et al. 1979; Alloin 
et al. 1979). The empirical oxygen abundance indicator R$_{23}$ = 
([OII]3727,3729 + [OIII]4959,5007)/H$_{\beta}$, suggested by Pagel et al. (1979), 
has found widespread acceptance and use for the oxygen abundance determination 
in H\,{\sc ii} regions where the temperature-sensitive lines are undetectable. This 
method will be referred to as the R$_{23}$ - method. Using the R$_{23}$ - method,
the characteristic oxygen abundances (the oxygen abundance at a predetermined 
galactocentric distance) and radial oxygen abundance gradients were obtained for 
a large sample of spiral galaxies (Vila-Costas and Edmunds (1992), Zaritsky et al. 
(1994), van Zee et al. (1998), among others).

Hovewer, the basic problem whether R$_{23}$ is an accurate abundance indicator 
is open for discussion (Zaritsky 1992, Kinkel and Rosa 1994, among others).
It has been found (Pilyugin 2000 (Paper I)) that the error in the oxygen 
abundance derived with the R$_{23}$ -- method involves two parts: the first is 
a random error and the second is a systematic error depending on the excitation 
parameter. A new way of oxygen abundance determination in H\,{\sc ii} regions 
(P -- method) was suggested (Paper I, Pilyugin 2001 (Paper II)). 
By comparing oxygen abundances in high-metallicity H\,{\sc ii} regions derived 
with the $T_{e}$ -- method and those derived with the P -- method, it was found 
that the precision of oxygen abundance determination with the P -- method is 
comparable to that of the $T_{e}$ -- method (Paper II). It was shown that the 
R$_{23}$ -- method provides more or less realistic oxygen abundances in 
high-excitation H\,{\sc ii} regions, but yields overestimates in 
low-excitation ones. Taking into account this fact together with the fact
known for a long time (Searle 1971, Smith 1975) that galaxies can show  
strong radial excitation gradients, in the sense that only low-excitation 
H\,{\sc ii} regions populate the central parts of some galaxies, one can expect 
that the central oxygen abundances and gradient slopes based on the 
(O/H)$_{R_{23}}$ data can be appreciably overestimated. 

This speculation can have far-reaching implications (the empirical estimation of
the oxygen yield, correlations of the behaviour of oxygen abundances with other
properties of spiral galaxies, etc). This speculation can be verified by comparison 
of the radial (O/H)$_{R_{23}}$ abundance distribution with the radial 
(O/H)$_{T_{e}}$ abundance distribution. The well-observed spiral galaxy M101 
provides such possibility. 

The comparison between radial distributions of (O/H)$_{T_{e}}$, (O/H)$_{P}$,
and (O/H)$_{R_{23}}$ abundances across the disk of M101 aiming to test the
credibility of the (O/H)$_{P}$ and (O/H)$_{R_{23}}$ abundances is a goal of 
the present study.

\section{The oxygen abundance distribution in M101}

\subsection{General comments}

Despite the fact that the spectroscopic data with detections of diagnostic 
emission lines in H\,{\sc ii} regions makes it possible to determine the accurate 
oxygen abundance (O/H)$_{T_{e}}$, 
the oxygen abundances in the same H\,{\sc ii} region with measured line ratios 
[OIII]$\lambda \lambda 4959, 5007 / \lambda 4363$ derived in different works can 
differ for three reasons: atomic data adopted, interpretation of the temperature 
structure (single characteristic $T_{e}$, two-zone model for $T_{e}$, model with 
small-scale temperature fluctuations) and errors in the line intensity 
measurements. Therefore the compilation of H\,{\sc ii} regions with original oxygen 
abundance determinations through the T$_{e}$ -- method from different works 
carried out over more than twenty years is not a set of homogeneous 
determinations. Accordingly, the available published spectra of H\,{\sc ii} regions with 
measured line ratios [OIII]$\lambda \lambda 4959, 5007 / \lambda 4363$ 
have been reanalysed to produce a homogeneous set. Two-zone models of H\,{\sc ii} 
regions with the algorithm for oxygen abundance determination from Pagel et al. 
(1992) and T$_{e}$([OII]) -- T$_{e}$([OIII]) relation from Garnett (1992) were 
adopted here. 

The determination of the (O/H)$_{P}$ oxygen abundances in high-metallicity H\,{\sc ii} 
regions has been considered in Paper II. The following expression has been 
suggested
\begin{equation}
12+log(O/H)_{P} = \frac{R_{23} + 54.2  + 59.45 P + 7.31 P^{2}}
                       {6.07  + 6.71 P + 0.37 P^{2} + 0.243 R_{23}}  ,
\label{equation:ohp}
\end{equation}
where $R_{23}$ =$R_{2}$ + $R_{3}$, 
$R_{2}$ = $I_{[OII] \lambda 3727+ \lambda 3729} /I_{H\beta }$, 
$R_{3}$ = $I_{[OIII] \lambda 4959+ \lambda 5007} /I_{H\beta }$, 
and P = $R_{3}$/$R_{23}$. 
Eq.(\ref{equation:ohp}) has been used here for the determination of the 
(O/H)$_{P}$ oxygen abundances.

Several workers have suggested calibrations of the R$_{23}$ in terms of the 
oxygen abundance (Edmunds and Pagel 1984, McCall et al. 1985; Dopita and Evans 
1986,  Zaritsky et al. 1994, among others). The most frequently used calibration
after Edmunds and Pagel (1984) has been adopted here for the determination of 
the (O/H)$_{R_{23}}$ oxygen abundances.

\subsection{Radial oxygen abundance gradient}

The nearby Sc galaxy M101 = NGC5457 has long served as the prototype system
for studying the radial oxygen abundance gradients in disks. Spectroscopic 
observations of H\,{\sc ii} regions in M101 have been carried out by many investigators
(Smith 1975; Shields and Searle 1978; Rayo, Peimbert, and Torres-Peimbert 1982;
McCall, Rybski, and Shields 1985; Torres-Peimbert, Peimbert, and Fierro 1989; 
Garnett and Kennicutt 1994; Kinkel and Rosa 1994; Kennicutt and Garnett 1996; 
van Zee et al. 1998; Garnett et al. 1999).

\begin{table}
\caption[]{\label{table:m101}
Oxygen abundances in H\,{\sc ii} regions of spiral galaxy M101.
The name of H\,{\sc ii} region is reported in column 1. The oxygen abundance computed
here in common way (through the T$_{e}$ -- method) is reported in column 2. 
The oxygen abundances in H\,{\sc ii} region Searle 5 derived with T$_{e}$([OII) and 
T$_{e}$([NII) are labeled with letters a and b respectively.
Source of line intensities measurements is given in column 3. 
The fractional radius $\rho$, normalized to the disk isophotal radius, 
is listed in column 4. 
}
\begin{center}
\begin{tabular}{lclc} \hline \hline
           &                     &            &                  \\  
H\,{\sc ii} region &12+log(O/H)$_{T_{e}}$& reference  &    $\rho$        \\  
           &                     &            &                  \\   \hline
Searle 5   &      8.55$^{a}$     &  KR94      &    0.22          \\   
           &      8.77$^{b}$     &  KR94      &                  \\   
+252-107   &      8.55           &  M85       &    0.33          \\   
NGC5461    &      8.40           &  S75       &    0.34          \\   
           &      8.50           &  R82       &                  \\   
           &      8.42           &  T89       &                  \\   
NGC5455    &      8.40           &  S75       &    0.48          \\   
           &      8.52           &  SS78      &                  \\   
           &      8.43           &  T89       &                  \\   
-347+276   &      8.45           &  vZ98      &    0.54          \\   
-459-053   &      8.32           &  vZ98      &    0.55          \\   
NGC5447    &      8.34           &  S75       &    0.55          \\   
Searle 12  &      8.19           &  S75       &    0.67          \\   
-398-436   &      8.06           &  vZ98      &    0.68          \\   
NGC5471    &      7.99           &  S75       &    0.84          \\   
           &      8.17           &  SS78      &                  \\   
           &      8.18           &  R82       &                  \\   
           &      8.11           &  T89       &                  \\   
           &      8.10           &  G99       &                  \\   
H681       &      7.91           &  GK94      &    1.04          \\   
+010+885   &      7.96           &  vZ98      &    1.04          \\   \hline  \hline 
\end{tabular}
\end{center}

\vspace{0.05cm}

{\it List of references}:
        
G99    --  Garnett, Shields, Peimbert, Torres-Peimbert,
           Skillman, Dufour, Terlevich E, Terlevich R, 1999
GK94   --  Garnett, Kennicutt, 1994;           
KR94   --  Kinkel, Rosa, 1994;
M85    --  McCall, Rybski, Shields, 1985;
R82    --  Rayo, Peimbert, Torres-Peimbert, 1982;
S75    --  Smith, 1975;
SS78   --  Shields, Searle, 1978;
T89    --  Torres-Peimbert, Peimbert, Fierro, 1989;
vZ98   --  van Zee, Salzer, Haynes, O'Donoghue, Balonek, 1998
\end{table}

The H\,{\sc ii} regions of M101 with  measured temperature-sensitive line ratios are 
listed in Table \ref{table:m101}. The name of H\,{\sc ii} region is reported in column 1. 
The oxygen abundance (O/H)$_{T_{e}}$ recomputed here is reported in column 2. 
Source of line intensities measurements is given in column 3. 
The fractional radius $\rho$, normalized to the disk isophotal radius
$\rho_{0}$, is listed in column 4. 
The galactocentric distances were taken from Kennicutt and Garnett (1996). 
The electron temperatures T$_{e}$([OIII]) in H\,{\sc ii} regions (with one exception,
H\,{\sc ii} region Searle 5) have been determined from the measurements of
[OIII]$\lambda \lambda 4959, 5007 / \lambda 4363$ line ratios, and 
the electron temperatures T$_{e}$([OII]) in H\,{\sc ii} regions have been derived from
the T$_{e}$([OII]) -- T$_{e}$([OIII]) relation of Garnett (1992). In the case
of H\,{\sc ii} region Searle 5 the electron temperature T$_{e}$([OIII]) cannot be
directly determined from observational data since the measurement of
[OIII]$\lambda \lambda 4959, 5007 / \lambda 4363$ line ratio is not available. 
Instead the temperature-sensitive lines [OII]$\lambda \lambda 7320, 7330$ and
[NII]$\lambda 5755$ were detected in deep spectrophotometry of H\,{\sc ii} region
Searle 5 (Kinkel and Rosa 1994). Then the electron temperature T$_{e}$([OIII]) 
in H\,{\sc ii} region Searle 5 has been derived from the T$_{e}$([OII]) -- T$_{e}$([OIII]) 
relation of Garnett (1992) using the value of T$_{e}$([OII]) reported by
Kinkel and Rosa (1994). The oxygen abundance in H\,{\sc ii} region Searle 5 based on 
the T$_{e}$([OII) is labeled with letter {\it a} in Table \ref{table:m101}.
The other value of oxygen abundance in H\,{\sc ii} region Searle 5 has been computed
using the value of T$_{e}$([NII) from Kinkel and Rosa (1994) and assuming 
T$_{e}$([OII]) = T$_{e}$([NII]).
The oxygen abundance in H\,{\sc ii} region Searle 5 based on the 
T$_{e}$([NII) is labeled with letter {\it b} in Table \ref{table:m101}.

\begin{figure}
\resizebox{1.00\hsize}{!}{\includegraphics[angle=0]{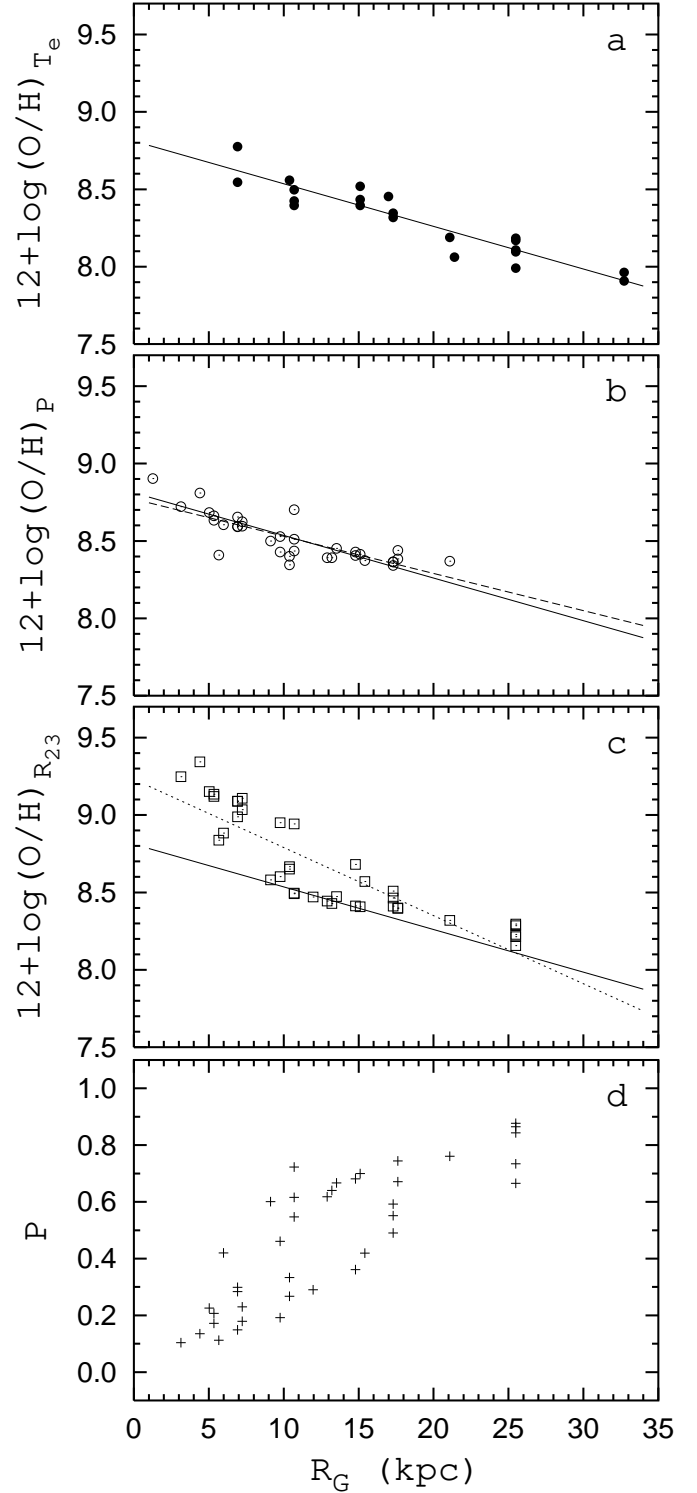}}
\caption{
Gradients in the properties of M101. 
{\bf a)} The (O/H)$_{T_{e}}$ abundance distribution (filled circles) and
the best fit (solid line).
{\bf b)} The (O/H)$_{P}$ abundance distribution (open circles) and
the best fit (dashed line). The solid line is the (O/H)$_{T_{e}}$ --  R$_G$ 
relation. 
{\bf c)} The (O/H)$_{R_{23}}$ abundance distribution (open squares) and
the best fit (dotted line). The solid line is the (O/H)$_{T_{e}}$ --  R$_G$ 
relation.
{\bf d)} The excitation parameter P as a function of galactocentric distance.
}
\label{figure:m101}
\end{figure}

In Fig.\ref{figure:m101}a we show oxygen abundances (O/H)$_{T_{e}}$ (the 
filled circles) for H\,{\sc ii} regions from Table \ref{table:m101} as a function of 
galactocentric distance. As seen in Fig.\ref{figure:m101}a the 
(O/H)$_{T_{e}}$ data is sufficient in quantity and quality for an accurate 
determination of the value of the oxygen abundance gradient within the M101. 
It is evident from Fig.\ref{figure:m101}a that the radial distribution of 
oxygen abundance within the disk of M101 can be reproduced by a single line. 
The best fit to the (O/H)$_{T_{e}}$ data is
\begin{equation}
12+log(O/H)_{T_{e}} = 8.81(\pm0.08) - 0.028(\pm0.005) \; R_{G}, 
\label{equation:m101te}
\end{equation}
where R$_{G}$ is in kpc. The (O/H)$_{T_{e}}$ -- R$_G$ relation is presented by the 
solid line in Fig.\ref{figure:m101}. The uncertainty of the value of central 
oxygen abundance is spesified by the mean deviation of positions of H\,{\sc ii} 
regions from the O/H -- R$_G$ relation. The maximum value of the error in the 
slope expressed in units of dex/$\rho_0$ (where $\rho_0$ is the isophotal radius) 
can be adopted equal to twice value of the mean deviation. The error in the 
slope expressed in units of dex/kpc can be estimated through the division of 
the error in the slope expressed in units of dex/$\rho_0$ by isophotal radius 
in kpc, that gives --0.028$\pm$0.005 dex/kpc. 

Fig.\ref{figure:m101}b shows the radial (O/H)$_{P}$ abundance distribution  
(the open circles) for H\,{\sc ii} regions from Kennicutt and Garnett (1996). 
The oxygen abundances in H\,{\sc ii} regions with galactocentric distances larger than 
$\sim$ 22 kpc are expected (Eq.\ref{equation:m101te}) to be less than 
12+logO/H=8.2, i.e they do not belong to the upper branch of the R$_{23}$ -- O/H 
diagram. Since the Eq.(\ref{equation:ohp}) can be used for oxygen abundance 
determination in H\,{\sc ii} regions of the upper branch of the R$_{23}$ -- O/H diagram 
only, the H\,{\sc ii} regions of M101 with galactocentric distances larger than 22 kpc 
are not presented in Fig.\ref{figure:m101}b. The best fit to the (O/H)$_{P}$ 
data is
\begin{equation}
12+log(O/H)_{P} = 8.76(\pm0.08) - 0.024(\pm0.005) \; R_{G}.
\label{equation:m101p}
\end{equation}
This (O/H)$_{P}$ -- R$_G$ relation is shown by the dashed line in the 
Fig.\ref{figure:m101}b. The mean residual from the (O/H)$_{P}$ -- R$_G$ relation 
is 0.084 dex. The corresponding error in the slope is $\pm$0.005 dex/kpc, 
i.e. the same as in the case of gradient traced by the (O/H)$_{T_{e}}$ abundances.
Inspection of Fig.\ref{figure:m101}b shows that the radial (O/H)$_{P}$ abundance 
distribution is in agreement with the (O/H)$_{T_{e}}$ distribution.

Fig.\ref{figure:m101}c shows the radial (O/H)$_{R_{23}}$ abundance distribution 
(the open squares) for the same H\,{\sc ii} regions as in Fig.\ref{figure:m101}b. 
The best fit to the (O/H)$_{R_{23}}$ data is
\begin{equation}
12+log(O/H)_{R_{23}} = 9.23(\pm0.16) - 0.044(\pm0.010) \; R_{G}.
\label{equation:m101r23}
\end{equation}
This (O/H)$_{R_{23}}$ -- R$_G$ relation is shown by the dotted line in 
Fig.\ref{figure:m101}c. The mean residual from the (O/H)$_{R_{23}}$ -- R$_G$ 
relation is 0.16 dex. The corresponding error in 
the slope is $\pm$0.010 dex/kpc. Inspection of Fig.\ref{figure:m101}c shows 
that the (O/H)$_{R_{23}}$ abundances in H\,{\sc ii} regions with large galactocentric 
distances are close to the (O/H)$_{T_{e}}$ -- R$_{G}$ relation, while the 
(O/H)$_{R_{23}}$ abundances in H\,{\sc ii} regions in the inner part of M101 have 
significant deviations from the (O/H)$_{T_{e}}$ -- R$_{G}$ relation. This 
behaviour of (O/H)$_{R_{23}}$ abundances with galactocentric distance confirms 
the conclusion of Paper II that the R$_{23}$ calibration of Edmunds and Pagel 
(1984) provides a realistic oxygen abundances in high-excitation H\,{\sc ii} 
regions, but overestimates them in low-excitation ones. Indeed,
the H\,{\sc ii} regions with galactocentric distances larger than around 17 kpc are 
high- and moderate-excitation (P values are higher than around 0.6) ones, 
Fig.\ref{figure:m101}d, their (O/H)$_{R_{23}}$ abundances are 
close to the (O/H)$_{T_{e}}$ -- R$_{G}$ relation. By contrast, the H\,{\sc ii} 
regions with galactocentric distances smaller than around 9 kpc are 
low-excitation (P values are less than around 0.3) ones, Fig.\ref{figure:m101}d, 
their (O/H)$_{R_{23}}$ abundances have significant deviations from the 
(O/H)$_{T_{e}}$ -- R$_{G}$ relation. 

The form of oxygen abundance gradient in M101 has been investigated in a number 
of studies (Evans 1986, Vila-Costas and Edmunds 1992, Henry and Howard 1995, 
Kennicutt and Garnett 1996). It has been noted by Kennicutt and 
Garnett (1996) and Roy and Walsh (1997) that the shape of the (O/H)$_{R_{23}}$ 
distribution is very sensitive to the precise form of the R$_{23}$ calibration. 
In the case of M101 Kennicutt and Garnett (1996) have found that the Edmunds 
and Pagel (1984) calibration produces a "steep -- shallow" break in the slope 
of the distribution (see also Vila-Costas and Edmunds 1992), 
 while the Dopita and Evans (1986) and McCall et al. (1985) 
calibrations yield distributions which are very well fitted with a single 
exponential function, though the slope of the exponential differs considerably 
between the two calibrations. A careful examination of Fig.\ref{figure:m101}
suggests that the distributions of the (O/H)$_{T_{e}}$, Fig.\ref{figure:m101}a, 
and (O/H)$_{P}$ abundances, Fig.\ref{figure:m101}b, can be fitted quite 
comfortably within the scatter by the same exponential with constant slope. 
There is hint of a turnover in the slope of the (O/H)$_{R_{23}}$ abundance 
distribution, Fig.\ref{figure:m101}c. This turnover is artifact and reflects 
the increase of error in (O/H)$_{R_{23}}$ abundance with decrease of 
galactocentric distance caused by variations in excitation parameter, 
Fig.\ref{figure:m101}d. 

An important result of the present study is the rather low value of the central 
oxygen abundance in M101. This is in agreement with the result of Kinkel \& Rosa 
(1994), who showed the need to lowering all H\,{\sc ii} region abundances 
obtained on the basis of the R$_{23}$ calibration after Edmunds and Pagel (1984) 
by at least 0.2 at intrinsic solar like O/H values and above.
One more large spiral galaxy in which the 
(O/H)$_{T_{e}}$ abundance distribution has been established is the Milky Way 
Galaxy. Caplan et al (2000) and Deharveng et al (2000) have analysed Galactic 
H\,{\sc ii} regions and have obtained the slope --0.0395 dex/kpc with central oxygen 
abundance 12+log(O/H) = 8.82, close to the solar value 12+log(O/H)$_{\odot}$ = 
8.83 (Grevesse and Sauval 1998), and 12+log(O/H) = 8.48 at the solar 
galactocentric distance, around a factor of 2 lower than the solar abundance.
Rodriques (1999) has considered seven bright Galactic H\,{\sc ii} regions with 
galactocentric distances in the range 6 -- 10 kpc and has found that all the
H\,{\sc ii} regions studied are characterized by similar abundances, 12+log(O/H) $\sim$ 
8.45$\pm$0.1. This value of the oxygen abundance at the solar radius is close to the value of 
the interstellar oxygen in the vicinity of the Sun which is about two-thirds of 
the solar oxygen abundance (Meyer et al. 1998). Those data taken together
suggest that the oxygen abundance at the solar radius is 1/2 $\div$ 2/3
of the solar oxygen abundance and increases up to about solar value 
in the centre of our Galaxy. The central oxygen abundance in M101 based on the 
(O/H)$_{T_{e}}$ (or (O/H)$_{P}$) abundances is close to that in our Galaxy. 

Thus, the consideration of M101 suggests the following. The available 
(O/H)$_{T_{e}}$ abundances allow to establish quite firmly the parameters of the 
radial oxygen abundance distribution (the central oxygen abundance and the 
gradient) within M101. The parameters of the (O/H)$_{P}$ abundance distribution 
are close to those of the (O/H)$_{T_{e}}$ abundance distribution. The parameters
of the (O/H)$_{R_{23}}$ abundance distribution differ significantly from those
of the (O/H)$_{T_{e}}$ abundance distribution: the values of the gradient and 
the central oxygen abundance based on the (O/H)$_{R_{23}}$ data are overestimated 
as compared to values derived from the (O/H)$_{T_{e}}$ distribution.

\subsection{The dispersion in abundance}

The dispersion in (O/H)$_{P}$ abundance at fixed radius is coincident with that 
in (O/H)$_{T_{e}}$ abundance, $\sim$ 0.08 dex. In the general case, the error 
in line intensity measurements can make contribution to the scatter at fixed 
galactocentric distance. The precision of present-day determinations of the 
oxygen abundances in high-metallicity H\,{\sc ii} regions through the 
T$_{e}$ -- method seems to be about 0.1dex (Deharveng et al. 2000). 
The precision of oxygen abundance determination with the P -- method is 
comparable to that of the $T_{e}$ -- method (Paper II). Those facts 
taken together suggest none, or only marginal, scatter at a given galatocentric 
distance. However, Kennicutt and Garnett (1996) advocate that the quality 
of their specta is sufficiently high that observational errors contribute 
negligibly to the dispersion. If this is the case it cannot be excluded that 
some part of the dispersion in 
(O/H)$_{P}$ abundance reflects the actual deviations of oxygen abundances
in individual H\,{\sc ii} regions from the general (O/H)$_{P}$ -- R$_{G}$ trend. 

The dispersion in (O/H)$_{R_{23}}$ abundance at fixed radius, $\sim$ 0.16 dex, 
is appreciable larger than that in (O/H)$_{P}$ abundance, $\sim$ 0.08 dex. The
following interpretation of this fact can be suggested. 
The extra dispersion in (O/H)$_{R_{23}}$ abundance at fixed radius is an artifact
and reflects the dispersion in excitation parameter P.

\begin{figure}
\resizebox{\hsize}{!}{\includegraphics[angle=0]{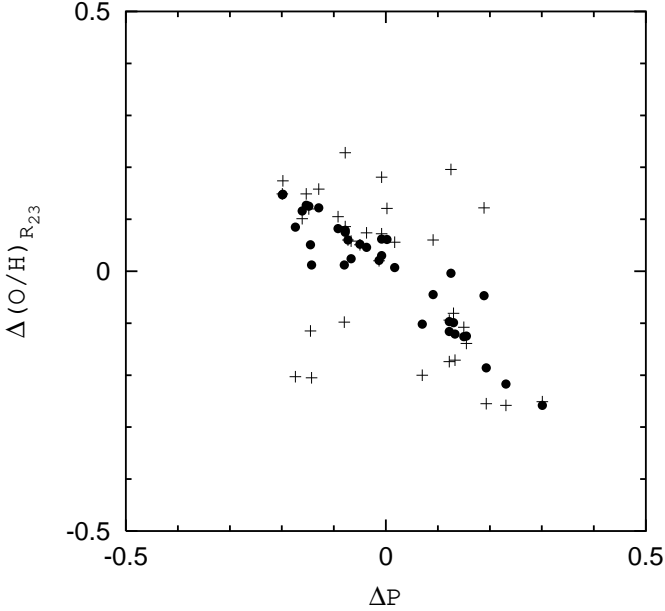}}
\caption{
The log(O/H)$_{R_{23}}$ residual from a mean linear fit, $\Delta$(O/H)$_{R_{23}}$, 
as a function of excitation parameter P residual, $\Delta$P, for H\,{\sc ii} regions 
from Kennicutt and Garnett (1996). The data obtained with original (O/H)$_{R_{23}}$ 
are shown by pluses, the data obtained with corrected (O/H)$_{R_{23}}$ are shown
by filled circles.
}
\label{figure:dpdz}
\end{figure}

Kennicutt and Garnett (1996) have also suggested that the dispersion 
in (O/H)$_{R_{23}}$ abundance at fixed radius can be attributed partly to variations 
in excitation parameter, but they did not find a firm confirmation of this 
suggestion. Following the strategy of Kennicutt and Garnett we fitted the 
radial variations in log(O/H)$_{R_{23}}$ and P by linear relations and 
considered correlation between the residuals in log(O/H)$_{R{23}}$ and P.
This correlation is shown in Fig.\ref{figure:dpdz} by pluses. As can seen
in Fig.\ref{figure:dpdz} the correlation between the residuals is very weak,
if at all. Why? We will demonstrate that this is  due to two reasons; {\it i)} 
abundance dispersion in the H\,{\sc ii} regions themselves, and 
{\it ii)} a feature of the sample.  

\begin{figure}
\resizebox{1.0\hsize}{!}{\includegraphics[angle=0]{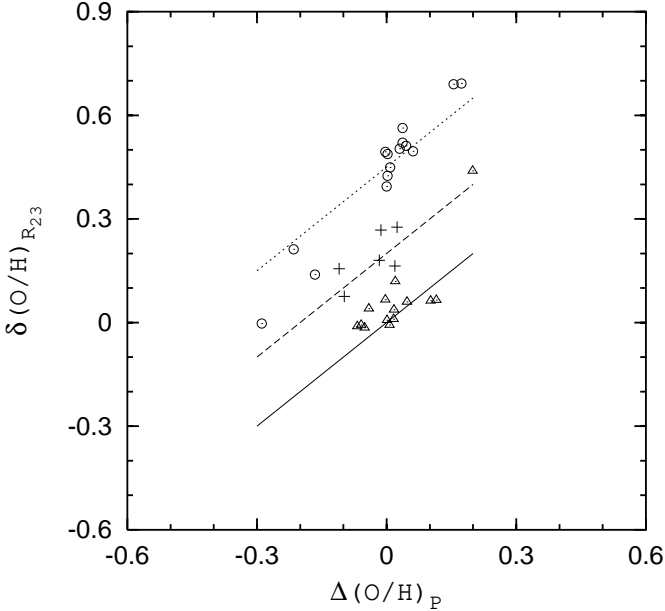}}
\caption{
The $\delta$(O/H)$_{R_{23}}$ is the deviation of log(O/H)$_{R_{23}}$ from the 
log(O/H)$_{P}$ -- R$_{G}$ fit. The $\Delta$(O/H)$_{P}$ is the log(O/H)$_{P}$ 
residual from the same fit. The triangles are H\,{\sc ii} regions with P$>$0.5,
the pluses are H\,{\sc ii} regions 0.5$>$P$>$0.3, the circles are those with 0.3$>$P.
The solid line is the relation $\delta$(O/H)$_{R_{23}}$ =  $\Delta$(O/H)$_{P}$, 
the dashed line is the relation $\delta$(O/H)$_{R_{23}}$ =  $\Delta$(O/H)$_{P}$
 + 0.2, the dotted line is the relation $\delta$(O/H)$_{R_{23}}$ = 
 $\Delta$(O/H)$_{P}$ +0.45.
}
\label{figure:dzdz}
\end{figure}

Fig.\ref{figure:dzdz} shows $\delta$(O/H)$_{R_{23}}$ (the deviation of individual 
log(O/H)$_{R_{23}}$ from log(O/H)$_{P}$ -- R$_{G}$ fit, Eq.\ref{equation:m101p})
versus $\Delta$(O/H)$_{P}$ (the log(O/H)$_{P}$ residual from the same fit)
for H\,{\sc ii} regions from 
Kennicutt and Garnett (1996). The triangles are H\,{\sc ii} regions with P$>$0.5,
the pluses are H\,{\sc ii} regions 0.5$>$P$>$0.3, the circles are those with 0.3$>$P.
Fig.\ref{figure:dzdz} shows that the variations in excitation parameter P 
(from $\sim$ 0.9 to $\sim$ 0.1, see Fig.\ref{figure:m101}d)
 result in the difference in (O/H)$_{R_{23}}$ as large 
as $\sim$ 0.6dex. 
Fig.\ref{figure:dzdz} shows that in the general case the $\delta$(O/H)$_{R_{23}}$ 
is the sum of two parts; the first is the $\Delta$(O/H)$_{P}$ and the second is
the deviation depending on the value of excitation parameter P.

The relevant feature of the sample of H\,{\sc ii} regions of M101 from Kennicutt and 
Garnett (1996) can be seen in Fig.\ref{figure:m101}d; the H\,{\sc ii} regions 
occupy a relatively narrow band in the P -- R$_{G}$ diagram. Due to this feature 
of the sample of H\,{\sc ii} regions the variations in excitation parameter P at 
fixed galactocentric distance do not exceed $\sim$ 0.4, or the maximum
deviation of excitation parameter from mean value is around $\pm$0.2. 
The maximum deviation around $\pm$0.15dex in (O/H)$_{R_{23}}$ corresponds to 
this maximum deviation of excitation parameter. 
Thus, the expected {\it maximum} deviation in (O/H)$_{R_{23}}$ due to the
deviation of excitation parameter from mean value is only twice the 
{\it average} actual deviation of oxygen abundances in individual H\,{\sc ii} 
regions from general (O/H)$_{P}$ -- R$_{G}$ trend, that can 
mask the  correlation between the residuals in log(O/H)$_{R{23}}$ and P.
Therefore the (O/H)$_{R_{23}}$ values were corrected for deviations caused by 
dispersion in the H\,{\sc ii} regions themselves
\begin{equation}
(O/H)^{c}_{R_{23}} = (O/H)_{R_{23}} - \Delta (O/H)_{P}.
\label{equation:dzdz}
\end{equation}
The $\Delta$P -- $\Delta$(O/H)$^{c}_{R_{23}}$ diagram for corrected
(O/H)$^{c}_{R_{23}}$ values has been constructed.
This diagram is shown in Fig.\ref{figure:dpdz} by filled circles. As can seen
in Fig.\ref{figure:dpdz} the correlation between the residuals is rather tight 
in this case. This is reliable confirmation that the dispersion 
in (O/H)$_{R_{23}}$ abundance at fixed radius is caused partly by variations 
in excitation parameter. 

\begin{figure}
\resizebox{\hsize}{!}{\includegraphics[angle=0]{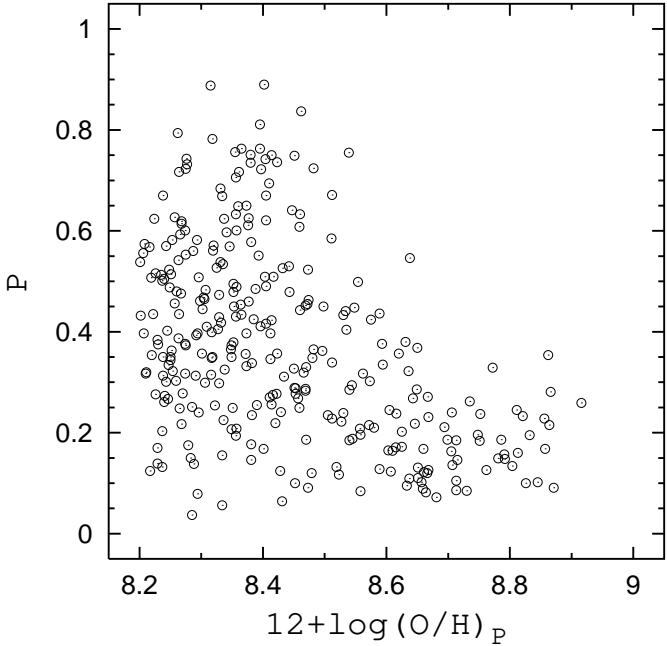}}
\caption{
The P -- (O/H)$_{P}$ diagram for H\,{\sc ii} regions from Zaritsky et al. (1994) and
van Zee et al. (1998). 
}
\label{figure:poh}
\end{figure}

As for the feature of Kennicutt and Garnett's sample of H\,{\sc ii} regions, 
the lack of high-excitation H\,{\sc ii} regions in central part of M101 can be 
explained by the ionization temperature gradient. The radial distribution of
the ionization temperature for the H\,{\sc ii} regions in M101 has been
investigated by Vilchez and Pagel (1988). They showed that there is a clear 
gradient in the temperature  along the disk of M101. The data for other
galaxies confirm the softening of the ionizing spectra with increasing metal 
abundance (Kennicutt et al. 2000, 
and references therein). Fig.\ref{figure:poh} shows the P -- (O/H)$_{P}$ 
diagram for a large sample of H\,{\sc ii} regions from Zaritsky et al. (1994) and
van Zee et al. (1998). Examination of Fig.\ref{figure:poh} shows that 
the maximum value of the excitation parameter P is dependent on the metallicity. 
Since excitation parameter P is an indicator of hardness of the ionizing 
radiation, the P -- (O/H)$_{P}$ diagram for large sample of H\,{\sc ii} regions 
confirms the softening of the ionizing spectra with increasing metal abundance.
But the lack of low-excitation H\,{\sc ii} at outer part of M101 seems to be conditioned 
by selection. 

\begin{figure}
\resizebox{\hsize}{!}{\includegraphics[angle=0]{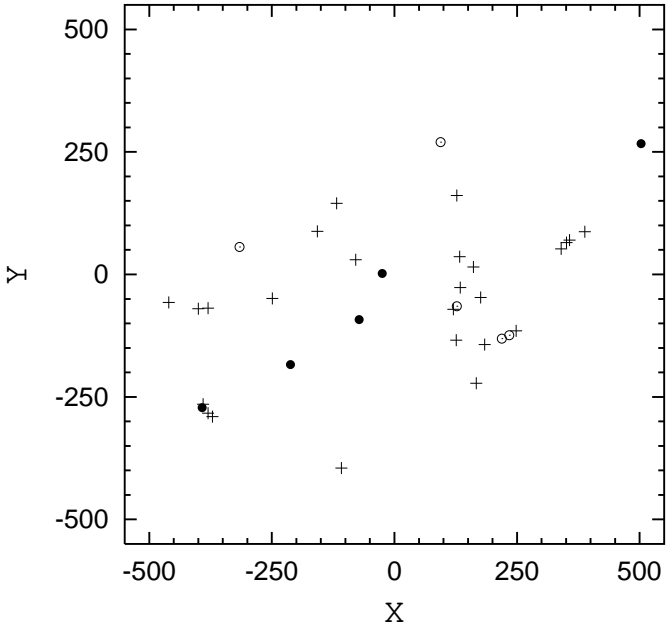}}
\caption{
The spatial distribution of the abundance residuals. Pluses are H\,{\sc ii} regions 
with -0.08 $<$ $\Delta$(O/H)$_{P}$ $<$ 0.08, filled circles are those with 
$\Delta$(O/H)$_{P}$ $>$ 0.08, open circles are those with $\Delta$(O/H)$_{P}$ $<$ 
-0.08.
}
\label{figure:xy}
\end{figure}

Kennicutt and Garnett (1996) concluded that there is hint that some of the 
dispersion in (O/H)$_{R_{23}}$ abundances in M101 is the result of a 
large-scale deviation from azimuthal symmetry. Is there spatial asymmetry in 
(O/H)$_{P}$  abundance residuals? Fig.\ref{figure:xy} shows the spatial 
distribution of (O/H)$_{P}$ abundance residuals. Pluses are H\,{\sc ii} regions 
with -0.08 $<$ $\Delta$(O/H)$_{P}$ $<$ 0.08, filled circles are those with 
$\Delta$(O/H)$_{P}$ $>$ 0.08, open circles are those with $\Delta$(O/H)$_{P}$ $<$ 
-0.08. Indeed, some asymmetry in the spatial distribution of (O/H)$_{P}$  
abundance residuals can be seen in Fig.\ref{figure:xy}. However, following 
Kennicutt and Garnett (1996) we can conclude that more data are needed to test 
whether this asymmetry is real.

\section{Conclusions}

The radial distributions of the oxygen abundances determined in three different 
ways (with the classic T$_{e}$ -- method, with the R$_{23}$ -- method, and with 
the P -- method) in H\,{\sc ii} regions of large spiral galaxy M101 have been compared. 

It has been found that the available (O/H)$_{T_{e}}$ abundances are sufficient 
in quantity and quality for an accurate determination of the parameters of the 
radial abundance distribution (the central oxygen abundance and the gradient).
We found that 12+log(O/H)$_{T_{e}}$ = 8.81($\pm$0.08) -- 
0.028($\pm$0.005)R$_{G}$(kpc).

It has been found that the parameters of the radial (O/H)$_{P}$ abundance 
distribution are close to those of the (O/H)$_{T_{e}}$ abundance distribution. 
We obtained that 
12+log(O/H)$_{P}$ = 8.76($\pm$0.08) -- 0.024($\pm$0.005)R$_{G}$(kpc).
This confirms the conclusion of Paper II that the (O/H)$_{P}$ abundances are as 
credible as the (O/H)$_{T_{e}}$ abundances. 

It has been obtained that the parameters of the radial O/H$_{R_{23}}$ abundance 
distribution differ significantly from those of the O/H$_{T_{e}}$ abundance 
distribution. We found that 12+log(O/H)$_{R_{23}}$ = 9.23($\pm$0.16) 
-- 0.044($\pm$0.010)R$_{G}$(kpc). This confirms our speculation that the central 
oxygen abundance and gradient slope based on the (O/H)$_{R_{23}}$ data can be 
appreciably overestimated. 

The dispersion in (O/H)$_{R_{23}}$ abundance at fixed radius is appreciable 
larger than that in (O/H)$_{P}$ abundance. It has been demonstrated that 
the extra dispersion in (O/H)$_{R_{23}}$ abundance is an artifact and reflects the 
dispersion in excitation parameter P.

\begin{acknowledgements}
It is a pleasure to thank J.M.~Vilchez and D.R.~Garnett for their helpful 
comments on this work. I thank the referee, Prof. B.E.J.~Pagel, for helpful c
omments and suggestions as well as improving the English text. This study was 
partly supported by the NATO grant PST.CLG.976036 and the Joint Research Project 
between Eastern Europe and Switzerland (SCOPE) No. 7UKPJ62178.
\end{acknowledgements}

\end{document}